\documentclass[aps,showpacs,nofootinbib]{revtex4}
\usepackage{epsf,latexsym}
\usepackage{amsfonts}
\usepackage{graphicx}
\usepackage[inline]{trackchanges}

\begin{document}

\title{Quantum states for a minimum-length spacetime}
\author{Alessandro Pesci\footnotetext{e-mail: pesci@bo.infn.it}}
\affiliation
{INFN Bologna, Via Irnerio 46, I-40126 Bologna, Italy}

\begin{abstract}
Starting from some results regarding 
the form of the Ricci scalar at a point $P$
in a (particle-like) spacetime endowed with a minimum distance,
we investigate how they might be accommodated,
specifically for the case of null separations,
in a as-simple-as-possible quantum structure
for spacetime at $P$,
and we try to accomplish this in terms 
of potentially operationally-defined concepts.  
In so doing, we provide a possible explicit form
for the operator expressing the Ricci scalar as a quantum
observable,
and give quantum-informational support,
thus regardless of or before field equations,
to associating with a patch of horizon
an entropy proportional to its area.
\end{abstract}


\maketitle

$ $
\section{Introduction}


From consideration of the combination of gravity
and quantum mechanics several results
from a variety of approaches have pointed
to the existence of a lower-limit length
\cite{DesA, Kadyshevsky1, Mea, DeWA_bis, BekC-02, Kadyshevsky2, DeWC, 
PadA-10, PadA-09, VenA} 
(\cite{GarA, HosA} 
for further references).
In recent works 
an intrinsic discreteness or particle nature
for spacetime has been considered 
by endowing it with a finite
lower-limit length $L_0$,
built 
\cite{KotE, Pad01, KotI}
in terms of a bitensor 
also called qmetric 
effectively embodying this.
Among other things, 
a result has been investigated
\cite{Pad01, KotI} 
(and \cite{PesP} for null separations) 
consisting in that 
the Ricci scalar $R_{(q)}$ in the qmetric at a point $P$
does not approach $R$ in the $L_0\to 0$ limit
and has the nature of kind of multivalued quantity,
somehow dependent on how we happen to probe 
spacetime curvature at $P$.
The persistence of the effect
in the $L_0\to 0$ limit,
is telling that the structure we get in the $L_0\to 0$ limit
(i.e., from a physical point of view, when the effects 
of a $L_0 \ne 0$ are undetectably small)
is not what we have with $L_0 = 0$ (i.e. ordinary spacetime).  

Such features,
far from being a mathematical accident of the model,
are found to unavoidably arise
once very basic conditions are met
(as the fact that for large separations the qmetric bitensor $q_{ab}$ 
provides distances which agree with what is foreseen
for ordinary metric,
and the well-posedness of the Green's function of the D'Alembertian
in the qmetric)
after input is taken from physics through the requirement of existence of
a non-vanishing lower-limit length.

We speculated then that
this might be accommodated by assuming
that embodying a non-vanishing $L_0$ goes hand in hand with requiring
an underlying (finite-dimensional) quantum structure
for spacetime at $P$,
a consistent description of which
would demand
on one side a constraint on the metric in the large scale
(in the form of field equations) \cite{PesQ},
and on another side the existence
of some, not better specified up to this stage, quantum operator
$\widehat{R}$, corresponding to the observable Ricci scalar,
with expectation value $R$ on the maximally mixed state \cite{PesR}.

The aim of present study
is to try to investigate further
this quantum space 
that such intrinsically-discrete 
or particle-like spacetime might
possess at $P$.
This, first of all trying to characterize its states,
and then finding an explicit expression
for the quantum operator $\widehat{R}$ acting on them.
We specifically restrict consideration to the qmetric
for null separated events,
as in \cite{PesQ} and \cite{PesR} on which we elaborate. 

Since the multivaluedness of $R_{(q)}$ mentioned above
comes about when trying to probe the Ricci scalar
reaching $P$ along different geodesics \cite{Pad01, KotI, PesP},
a further exploration of this fact seems to naturally
hint to operational notions/procedures.
This apparently suggests that
a proper framework
for further progress
might be
the consideration of gravity 
in an as-much-as-possible
operational setting, that is 
in terms of information acquired 
by physical bodies.

Much progress has taken place in recent times
concerning an operational characterization of gravity,
in particular regarding the formulation of the equivalence principle,
which is at the heart of general relativity, 
in quantum operational terms,
namely with quantum theory considered 
first of all as a means
of information processing.
This has been specifically addressed in
\cite{Hardy.1, Hardy.2} 
and in \cite{Gia.1},
with emphasis in this last work
on considering
quantum reference frames as something 
corresponding to 
real quantum systems (rather than
abstract quantum coordinates) 
as explicitly described in \cite{Gia.2}. 
On top of this, 
starting from \cite{BosA, MarlettoVedral}
several operational procedures have been devised,
which might allow in a not far future a direct
experimental verification of a quantumness,
or non-classicality at least, of the gravitational field,
or, quite on the contrary, to rule out any quantum superposition
of gravitational fields, along what envisaged in \cite{Penrose}.

Also, the strong case has been made that
the quantum space which ought
to describe gravity is locally finite-dimensional
\cite{BaoCarrollSingh},
this essentially arising
from the Bekenstein bound \cite{BekD}.
On the other hand,
finite-dimensional quantum mechanics
has been shown to be entirely derivable
from a few axioms of general operational theories
\cite{Hardy.3, Hardy.4, D'Ariano.1, D'Ariano.2, D'Ariano.3}.

All this,
suggests that in the operational  methods
there might be all the ingredients
needed for a new, deeper-level
understanding of the quantum spacetime
at a point.
This paper is meant as a tentative step 
in this direction
building on
the results with qmetric.

\section{Quantum spacetime at a point: null states}\label{quantum.spacetime}

As mentioned,
spacetime endowed with a minimum-length $L_0$
exhibits in the small scale
a peculiar structure at any event $P$
related to the fact that 
the qmetric Ricci scalar $R_{(q)}$ does not tend 
to the ordinary Ricci scalar $R$
in the $L_0\to 0$ limit.
In particular, the limit value depends on the direction of approach 
to $P$.
This is puzzling because the Ricci scalar ought to be 
determined completely by the assigned spacetime 
when giving $P$ (with no need of further specifications at $P$);
to be sure this ought to be the case at least when $L_0\to 0$,
limit in which all dependence on the direction along which in the qmetric
we postulated the existence of an $L_0\ne 0$ goes to be lost.
In this limit one would expect $R_{(q)}$ to not depend on the direction of
approach and to definitely be $R$, 
but as a matter of fact this is not the case.
A crucial further feature observed in the Ricci scalar of the qmetric
is that if we average the obtained limiting values of $R_{(q)}$
over the possible orthogonal directions 
we do get $R$. 

It appears here sort of
similarity between the just described results
and what we would get if we went to
measure a quantum observable corresponding to
the Ricci scalar at (classical) event $P$
with probes reaching $P$ along orthogonal directions, 
each measurement possibly consisting in taking the expectation 
value of quantum Ricci
according to some suitable state.
The purpose of the paper is to try to investigate 
this similarity
and possibly turn it
into something more definite and workable on.
As we will see below,
this is done by introducing an abstract, 
finite-dimensional Hilbert space $H$ attached 
to (classical) event $P$,
capable of describing all the states of qmetric spacetime at $P$
as well as the observables at $P$,
specifically the Ricci scalar.

Having this finite-dim Hilbert space reproducing qmetric results
might be enough in that 
$d$-dimensional Hilbert spaces
with assigned finite $d$ are all equivalent to each other.
But, which is the physical system which these states 
are supposed to refer to?
It is generically ``the spacetime at $P$'' as probed by that congruence.
But this entails a reference body $A$ (the observer) and a test body $B$
reaching $P$ along the possible geodesics, both generically quantum.
We see on one hand minimum length results hint at a kind of 
microstructure for spacetime, on the other the probing
of this possible microstructure
means to resort to observers and test bodies.
This quite naturally leads to join the minimum-length results
with an operational view of gravity.   

%
Following Einstein himself,
we consider a point $P$ in spacetime
not merely as a mathematical entity
in an abstract manifold,
but as something physically defined by some
crossing or coincidence event
among material bodies.
In the same spirit, 
and following the emphasis
drawn on this for a long time now 
(see \cite{Aharonov}),
we consider as reference frames not merely
a set of local coordinates,
but actual bodies,
quantum bodies since we are going to a small scale,
in terms of which the motion of other quantum
bodies, 
the test particles, 
is described. 
Point $P$ is regarded as a coincidence event
involving a material reference frame (system $A$),
generically, but not necessarily, 
part of the matter which sources the 
gravitational field,
and a test particle (system $B$).
In general, we can assume 
that both the reference body
and the test particle give, if any,
only a slight perturbation of
the field;
this, in the perspective of performing measurements
disturbing as little as possible.
It is clear however that in the vacuum this cannot apply:
in this case, the material frame and the test particle
are non-negligible sources of the field 
which is measured at $P$.

In fact,
what we would like to do 
is to
explore this way
the short scales
(in principle, down to the Planck scale),
and going to very short scales
implies test and/or reference particles
of very high momentum,
whose own field
we can naively expect to definitely dominate at $P$
even in the presence of matter as a source.
This would apparently deprive of any
possibility of success the attempt
to measure in the short scale
the properties of a given configuration at $P$. 

We have to consider however that,
as we will see in more detail below,
the departure from
the classical result in the expression
of Ricci scalar in the qmetric,
is at leading order insensitive
to the value of $L_0$.
The physical effects we want to study in this work
are exactly this `something' which is present
due to an $L_0 \ne 0$, and yet stays unchanged
from an $L_0$ (relatively) large down
to an $L_0$ vanishingly small.
To try to investigate which might be the characteristics of
this `something', we {\it assume} we can operationally probe it;
this is what we are supposing in this work.
This amounts to assume that $L_0$, which is a free parameter
of the model, is (relatively) large.   
We have to bear in mind however
that with this
we are not saying that $L_0$ is necessarily (relatively) large:
we are only taking advantage of a not exceedingly small $L_0$
to study some possible consequences of an effect which is
scale-independent in the small scales.

How large might $L_0$ be taken?
We will see that,
concerning the qmetric taken alone,
a relevant reference scale is
$L_R = 1/\sqrt{R_{ab} l^a l^b}$
(with $R_{ab}$ the Ricci tensor and $l^a$ the
tangent to the geodesic under consideration), 
kind of curvature length scale
much larger than the Planck scale.
We can consider
scales $\ell \ll L_R$ for $L_0$ 
(but with still $\ell \gg L_{Pl}$)
and as test particle e.g. a photon with 
wavelength $l \sim \ell$,
such that the energy density $\rho_\gamma$
of the photon (in a cube of edge $\ell$)
is $\rho_\gamma \ll \rho_s$ where $\rho_s$
is the source at $P$ of the field.
This is for sure possible if e.g. the source
of gravitational field are massive particles.

We emphasize however that the configuration 
we are considering, even if with an operational flavor,
is in the form of a gedankenexperiment.
What we are using is a theoretically viable setup
to highlight some consequences
of the qmetric approach.
The actual feasibility of it is another story.
As for the latter,
the absence of any signal of a 
$L_0 \ne 0$ at Large Hadron Collider,
suggests indeed to take
$L_0 < 1/10 \, {\rm TeV}^{-1} = 2.0 \cdot 10^{-5} {\rm fm}$
(cf. {\cite{Nicolini}})
(which is in the intermediate scales $\ell$ above).
This means that as body $A$ we need mass-energies 
$m > 10 \,\, {\rm TeV}$ and we have to approach $A$ at $\approx \, {\rm fm}$
scales (thus $A$ has to be structureless at these energies).
To meet these conditions in a controlled way
we have to think to colliders,
and we can generically expect
that the theoretical phenomenon described here
might have chances of actual experimental scrutiny
as soon as signs of a $L_0 \ne 0$ become visible at colliders.

Summing up, what we do here
is --leaving for the moment the actual experimental verification apart--
to describe the spacetime at $P$ with the qmetric
with $L_0$ in the intermediate range mentioned above.
And
the physical system to which the Hilbert space above refers
(namely the spacetime at $P$)
ought to be the composite system made up of 
$A$ (material reference body or observer) and 
$B$ (test particle).
From basic principles of quantum physics 
\cite{Peres}
we have then
that this Hilbert space $H$ 
is the tensor product $A \otimes B$,
denoting for simplicity of notation 
with the same letter the state space 
and the system to which it refers in each case.

To proceed,
we have to specify the states of $A$ and $B$
and of $H = A \otimes B$.
To this aim 
let us first recall in some more detail 
the results obtained
through the qmetric
which we would take advantage of in guessing
for a quantum description.

As mentioned in the introduction,
we here consider the case of null separations.
This implies that,
in the consideration of the Ricci scalar
we are restricting attention to
spacetimes
(including reference body + test particle,
or from these two alone in case of the vacuum)
such that a congruence
of null geodesics in all spatial directions from $P$
has in it all what is needed
to fix the Ricci scalar $R$ at $P$
(in spite of being the congruence short of one dimension
as compared to the spacetime $M$) \cite{PesP}.
In these circumstances,
using a congruence of null geodesics emerging from $P$
parameterized by length according
to a local observer at $P$
which plays the role of reference system $A$,
from the relation ($M$ $D$-dimensional)

\begin{eqnarray}\label{Ricci}
R =
\sum_{i=1}^{D-1} R_{ab} \, {l^a}_i \, {l^b}_i,
\end{eqnarray}
for $L_0$ small we definitely get \cite{PesR}

\begin{eqnarray}\label{expectation_value}
\frac{1}{D-1} \, \sum_{i=1}^{D-1} R_{(q), \, {l^a}_i, L_0} 
&=& 
R
\nonumber \\
&=&
\frac{1}{D-1} \, \sum_{i=1}^{D-1} R_{(q), \, {l^a}_i}. 
\end{eqnarray}
Here,
$R_{ab}$ and $R \equiv g^{ab} R_{ab}$ are the (ordinary-metric)
Ricci tensor and Ricci scalar at $P$
(quantities denoted without index ${(q)}$ 
refer to ordinary metric);
${l^a}_i$ is the (null) tangent vector at $P$ to the geodesic $i$,
with the geodesics taken in spatial directions orthogonal to each other;
and \cite{PesP}
\begin{eqnarray}\label{Rq}
R_{(q), \, l^a, L_0} 
&=& 
(D-1) \, R_{ab} l^a l^b + {\cal O}\bigg(\frac{L_0}{L_R} \, R_{ab} l^a l^b\bigg)
\end{eqnarray}
is the qmetric Ricci scalar at $P$
as probed through geodesic with tangent $l^a$ at $P$
with $L_0$ being the minimum length which characterizes the qmetric 
spacetime
and the magnitude of high order terms taken from \cite{PesQ};
$R_{(q), \, l^a} \equiv \lim_{L_0\to 0} R_{(q), \, l^a, \, L_0}$.
We see that,
as anticipated above,
the leading term does not depend on $L_0$ 
and the high order terms are smaller
by a factor $L_0/L_R$, involving the length scale $L_R$.
As mentioned,
equation (\ref{Rq}) shows that a same entity,
the qmetric Ricci scalar at $P$,
happens to take different values
depending on through which geodesic we look at it;
and this might be interpreted \cite{PesR, PesQ}
as suggesting that
the qmetric spacetime at $P$ should be regarded
as a superposition of spacetimes, each with its own
metric, a characteristic non-classical feature.
As repeatedly noticed,
this phenomenon arises 
in a minimum-length $L_0 \ne 0$ description,
yet persists virtually unaltered 
in the $L_0 \to 0$ limit.
It clearly stays there 
when (reduced) Planck's constant $\hbar \to 0$
in case 
$L_0$ is something different
from the Planck length $L_{Pl}$ with  
$L_0 \ne 0$ when $\hbar \to 0$.
But this happens also 
even if we think 
of $L_0$ as vanishing with $\hbar$,
like e.g. if it is something proportional
to Planck length, 
$L_0  =  C \, L_{Pl}  =  C \, \sqrt{G \hbar/c^3}$ with  $C$ a constant
(and $G$ and $c$ Newton's constant and speed of light in vacuum), 
which gives $L_0\to 0$ when $\hbar \to 0$.
This non-classical structure of spacetime
is thus something that,
when spelled out in particular in quantum terms,
is in any case apparently not ${\cal O}(\hbar)$,
with the meaning that it is not vanishing 
when assuming a vanishing $\hbar$.
It looks like having a status akin
to Bell's inequalities. 
All this seems to resonate with the results \cite{BosA, BosB}
and \cite{MarlettoVedral},
specifically in the description provided in \cite{RovS_2}.

Relation (\ref{expectation_value})
exhibits the value of the classical Ricci scalar $R$
as an average over $D-1$ qmetric terms.
One way to look at this \cite{PesR},
is to take it
as suggesting reference to
a $(D-1)$-dimensional quantum space
(Hilbert space corresponding to
$D-1$ perfectly discriminable states),
with equation (\ref{expectation_value})
expressing the expectation value 
of the Ricci scalar, considered as a quantum observable,
on the maximally
mixed state.
We try here to bring 
this perspective
a little further.

Let us consider this quantum space 
as the $(D-1)$-dimensional Hilbert space $H = A \otimes B$
describing the quantum states
(of the system consisting of the spacetime at $P$)
associated to null directions at $P$.
That is,
rooted in the just considered similarities,
the assumptions in building $H$ are:
i) $H$ is finite dimensional, the dimension being that
of the (sub)manifold swept by the congruence of geodesics used to
probe spacetime around $P$ (this is $D-1$ where $D$ is the dimension
of spacetime, since we use a congruence of null geodesics from $P$); 
ii) we take (spatially) orthogonal directions as associated to perfectly
distinguishable states, 
thus to states orthogonal according to the internal  
product of the Hilbert space.
 
We then construct $H$ as follows.
%
Denoting
$L = \{k^a \ {\rm at} \ P: k^a \ {\rm null \ and \ future \ directed} \}$, 
we introduce a 
correspondence $f{:} \ k^a = (k^0, \vec{k})  \longmapsto |\vec{k}\rangle$
from $L$
to an abstract $(D-1)$-dimensional Hilbert space
with internal product $\langle\cdot|\cdot\rangle$.
For
${l^a}_i
= {e^a}_0 + {e^a}_i
= (1, \vec{e_i})$, $i = 1, ..., D-1$
(here and hereafter $i, j$ label vectors of the base;
${e^a}_0$ 
is unit timelike),
this gives
$
f({l^a}_i)
=
|\vec{e_i}\rangle
\equiv
|i\rangle.
$
We introduce $H$ as the abstract $(D-1)$-dimensional Hilbert space
with internal product $\langle \cdot | \cdot \rangle$,
obtained as the complex span of the elements 
$|e_i\rangle = f({l^a}_i)$, 
$i = 1, ..., D-1$,
these being coinceived as pure states 
defined to be orthonormal according 
to $\langle\cdot|\cdot\rangle$.
This implements the view
that, as for the system consisting of the coincidence at $P$
(i.e. of the composition of the reference body $A$ and
the test particle $B$)
the spatial direction associated to a vector ${l^a}_i$
is exactly definite, and is thus represented
by a pure state, i.e. by a vector in $H$. 

The bijective correspondence $f{:} \, L\to H $ 
is not a correspondence between vector spaces
($L$ is not a vector space).
This ought not to be a problem however,
provided no incongruences arise
in any linear operation in $L$
which maps $L$ to itself;
in this case,
what we have to require
is that
the corresponding operation in $H$
brings to a vector $\vec{k'}$ which is precisely the image
through $f$
of the vector $k'^a$ we have got in $L$
as a result of the operation.
This however is of course guaranteed by the fact
that the vector $\vec{k}$  is part of the vector $k^a$
and thus any linear operation on $k^a$ which
brings to a vector $k'^a \in L$
involves a linear operation 
which brings from $\vec{k}$
to exactly the component $\vec{k'}$
orthogonal to ${e^a}_0$
of the vector $k'^a$.
This happens in particular
for the case of scalar multiplication by 
$\eta \ge 0$,
and for rotations in the subspace 
orthogonal to ${e^a}_0$;
for both,
one easily verifies that the vector one gets in $H$
is the image of the vector we get in $L$, indeed

\begin{eqnarray}\label{xq95.1}
\eta f(k^a) 
= \eta |\vec{k}\rangle
= |\eta \vec{k}\rangle
= f(\eta k^a),
\end{eqnarray}
and

\begin{eqnarray}\label{xq95.2}
Q f(k^a) 
= Q |\vec{k}\rangle
= |Q \vec{k}\rangle
= f\big(\widetilde{Q} k^a\big) 
\end{eqnarray}
where
$\widetilde{Q} = (1, Q)$
with 
$Q$ a $(D-1) \times (D-1)$ orthogonal matrix
expressing a rotation in the subspace 
orthogonal to ${e^a}_0$.

\section{A quantum observable for the Ricci scalar}

In the previous section,
we established a correspondence
between null vectors $k^a$ of the local frame 
of coincidence event $P$
and vectors of the Hilbert space $H$,
with every element of $L$ represented in $H$
this way.
Our next task
is now to be able to describe the Ricci scalar
as a quantum observable, namely to express it 
in the form of a Hermitian
operator $\widehat R$ of $H$.

The results \cite{PesQ, PesR},
have hinted to that the quantity 
$R_{(q), \, l^a, L_0}$ in (\ref{Rq}),
or its $L_0 \to 0$ limit form
$R_{(q), \, l^a} = (D-1) R_{ab} l^a l^b$,
might be taken as the output we get from a probe of the Ricci scalar
through a geodesic with tangent $l^a$ at $P$,
this in turn tentatively giving kind of a yet to be precisely
defined
expectation value on the maximally mixed state
coinciding with the ordinary Ricci scalar $R$ at $P$.
This exhibits expression (\ref{expectation_value})
as a potential candidate from which 
to start trying to infer an expression
for $\widehat R$.  

Looking at it,
from what we did so far
it becomes quite natural to think of simply replacing
the null vectors ${l^a}_i$ in the terms
\begin{eqnarray}
R_{(q), \, {l^a}_i}
= (D-1) \, R_{ab} \, {l^a}_i \, {l^b}_i,
\end{eqnarray}
in (\ref{expectation_value}),
with
their quantum counterparts in $H$,
leaving $R_{ab}$ as it is.
This corresponds to the presumption
that the quantum nature ascribed 
to the (quantum) Ricci scalar
might be captured in the simplest manner
by resorting to the null quantum states
as replacing the null fields, 
as well as by the occurrence 
of the factor $(D-1)$.

Inspecting however the terms $R_{ab} \, {l^a}_i \, {l^b}_i$
or, more generally,
quantities of the kind 
$R_{ab} \, {l^a}_i \, {l^b}_j$,
we see that even in this as-simple-as-possible prescription 
we have to face the problem of how
to express the sums over indices $a$ and $b$ in terms
of vectors of $H$.
Essentially the problem is that of being able to manage
the time component ${l^0}_i$ of ${l^a}_i$,
that is terms of the kind
$R_{00} {l^0}_i {l^0}_j$ or
$R_{0\alpha} {l^0}_i {l^\alpha}_j$
($\alpha = 1, ..., D-1$, 
and in the expression we have 
implicit sum on the repeated index $\alpha$).

Since the vector ${l^a}_i = (1, \vec{e_i})$ gets mapped into
the state vector $|i\rangle$,
we do this by introducing
the symbol

\begin{eqnarray}\label{96.1}
|{l^a}_i\rangle 
\equiv
\big(|i\rangle, 0, 0, ..., |i\rangle, 0, ..., 0\big)
=
{l^a}_i \, |i\rangle
\end{eqnarray} 
(no sum on repeated $i$ implied).
It denotes
a string of $D$ (dimension of spacetime)
vectors of the Hilbert space.
Index $a$ selects the place in the string 
($a = 0, 1, ..., D-1$);
the first entry in the string, the time component in index $a$, 
has the state vector
defined by the remaining entries (the space components in index $a$)
(in (\ref{96.1}) then the same state vector $|i\rangle$ appears
both in place $a = 0$ and $a = i$).  
In general,
for $l^a = (1, \hat{k})$ null ($\hat{k}$ versor), 
we have

\begin{eqnarray}
|l^a\rangle 
\equiv
\big(|\vec{k}\rangle, k_1 |1\rangle, 
k_2 |2\rangle, ..., k_{D-1}|D-1\rangle\big).
\end{eqnarray}
Using this, 
we can write
the quantum observable representing the Ricci scalar
as

\begin{eqnarray}\label{97.4}
\widehat{R}
&\equiv&
(D-1) \sum_{i, j = 1}^{D-1}
R_{ab} |{l^a}_i\rangle\langle{l^b}_j| 
\nonumber \\ 
&=&
(D-1) \sum_{i, j = 1}^{D-1}
R_{ab} \, {l^a}_i \, {l^b}_j \, |i\rangle\langle j |.
\end{eqnarray}
Notice that the operator $\widehat{R}$ is real symmetric,
then Hermitian.

The cases of Ricci-flat and of Einstein spacetimes
are in view of this result somehow special or peculiar,
in that the curvature operator $\widehat R$ we get from ({\ref{97.4}})
is for them identically 0.
As a matter of fact the qmetric Ricci scalar turns out
to coincide with the ordinary Ricci scalar for Ricci-flat spacetimes
(and also for Einstein spacetimes in the case of the qmetric
based on null separations).
We notice however that,
while this surely deserves further understanding,
it seems to have little effect in a context in which the spacetime
is probed operationally with a reference body (and a test particle).
The stress-energy tensor of the reference body itself
generically guarantees indeed $R_{ab} \ne 0$ and $R_{ab} l^a l^b \ne 0$ 
at P
(both in case it is the only matter present and apparently
also if it is part of the source distribution
(an exception being the cosmological fluid,
namely interpreting the cosmological constant effects as due to
a fluid with $p = -\rho$)).

Calculating the expectation value of $\widehat{R}$
over the maximally mixed state
$\chi = \frac{1}{D-1} \sum_{i = 1}^{D-1} |i\rangle\langle i|$
of $H$,
we get

\begin{eqnarray}\label{97.7}
\langle \widehat{R} \rangle_\chi
&=&
\rm{tr}\big(\chi \, \widehat{R}\big)
\nonumber \\
&=&
\sum_{i = 1}^{D-1} \frac{1}{D-1} \, 
\rm{tr}\big(|i\rangle\langle i| \widehat{R}\big)
\nonumber \\
&=&
\sum_{i = 1}^{D-1} \frac{1}{D-1} \, 
\langle i| \widehat{R} | i \rangle
\nonumber \\
&=&
\sum_{i = 1}^{D-1} R_{ab} \, {l^a}_i \, {l^b}_i
\nonumber \\
&=&
R, 
\end{eqnarray}
where the second equality is from the linearity of the trace,
and the last from equation (\ref{Ricci}).
Clearly the expectation value is the same whichever is the basis
we can have chosen for $H$.
We mentioned that the multivaluedness of $R_{(q)}$,
with the value depending on the geodesic with which we reach $P$,
can be interpreted as suggesting that the spacetime
at coincidence $P$ can be interpreted as a superposition
of geometries. 
The value we obtain,
namely the ordinary Ricci scalar $R$ at $P$,
fits then with what one would expect 
from randomly probing the quantum Ricci scalar
with a flat distribution in direction.
The operator $\widehat{R}$ as defined by (\ref{97.4})
would be thus a possible explicit expression of a quantum observable
corresponding to the Ricci scalar
along the lines envisaged in \cite{PesR}.

Since $\widehat{R}$ is Hermitian on a finite dimensional Hilbert space,
from the spectral decomposition theorem 
(see e.g. \cite{NielsenChuang}) we know it is diagonalizable.
Being it real, this is accomplished by an orthogonal
matrix $Q$.
We have

\begin{eqnarray}\label{98.1}
\widehat{R} 
&=& 
\sum_{i'=1}^{D-1} \lambda_{i'} |i' \rangle\langle i'|
\end{eqnarray}
$i' = 1, ..., D-1$,
with

\begin{eqnarray}\label{98.2}
|i'\rangle = \sum_{i = 1}^{D-1} Q_{i' i} |i\rangle,
\end{eqnarray}
and correspondingly
\begin{eqnarray}\label{98.3}
\vec{e}_{i'} = \sum_{i = 1}^{D-1} Q_{i' i} \, \vec{e}_{i}.
\end{eqnarray}
The eigenvalues $\lambda_{i'}$ are given by 

\begin{eqnarray}\label{100.1}
\lambda_{i'}
&=&
\langle i' | \widehat{R} | i' \rangle
\nonumber \\
&=&
(D-1) \sum_{i, j = 1}^{D-1} R_{ab} \, {l^a}_i \, {l^b}_j \,
\langle i' | i \rangle \langle j | i' \rangle
\nonumber \\
&=&
(D-1) \sum_{i, j = 1}^{D-1} R_{ab} \, {l^a}_i \, {l^b}_j \,
\langle i | i' \rangle \langle j | i' \rangle
\nonumber \\
&=& 
(D-1) \sum_{i, j = 1}^{D-1} R_{ab} \, {l^a}_i \, {l^b}_j \, Q_{i' i} \, Q_{i' j}
\nonumber \\
&=&
(D-1) \, R_{ab} \, {l^a}_{i'} \, {l^b}_{i'}.
\end{eqnarray}
Here, 
the third equality comes from being 
$\langle i' | i \rangle$ real, 
which gives 
$\langle i' | i \rangle = \langle i | i' \rangle$;
the last from
\begin{eqnarray}\label{100.2}
\sum_{i=1}^{D-1} Q_{i' i} \, {l^a}_i 
=
\Big(1, \sum_{i=1}^{D-1} Q_{i' i} \, \vec{e}_i\Big)
=
(1, \vec{e}_{i'})
=
{l^a}_{i'}.
\end{eqnarray}
Equation (\ref{100.1}) exhibits
the $\lambda_{i'}$'s as the quantities
$R_{(q), \, {l^a}_{i'}}$
we find for geodesics with tangent ${l^a}_{i'}$ at $P$
such that ${l^a}_{i'}$ is ${l^a}_i$ rotated
by the matrix $(1, Q)$ with $Q$ that same matrix 
which describes the rotation from
the basis $\{|i\rangle\}$ to $\{|i'\rangle\}$.

Notice that, when diagonalizing $\widehat{R}$,
we change the basis in $H$ not the local frame
at $P$.
What happens in the tangent space at $P$ is that we move
from the vectors ${l^a}_i$ to the vectors ${l^a}_{i'}$.
Then, each single term $R_{ab} \, {l^a}_i \, {l^b}_i$ goes into
$R_{ab} \, {l^a}_{i'} \, {l^b}_{i'}$ and changes in this operation
(by contrast with what we would get were our operation
a change of local frame: 
$R_{ab} \, {l^a}_i \, {l^b}_i \mapsto 
R_{a' b'} \, {l^{a'}}_i \, {l^{b'}}_i = R_{ab} \, {l^a}_i \, {l^b}_i$).

Equation (\ref{98.1}) can be read as
$\widehat{R} = \sum_{i'=1}^{D-1} \lambda_{i'} P_{i'}$,
where the operator $P_{i'}$ is the projector onto
the (1-dim) eigenspace of $\widehat{R}$ with eigenvalue $\lambda_{i'}$.
These operators are orthogonal to each other 
and form a projected-value measure of the observable $\widehat{R}$. 
From basic tenets of quantum mechanics,
in a measurement of $\widehat{R}$
immediately after another one which gave 
as a result $\lambda_{i'}$, we have to find again
$\lambda_{i'}$ with certainty.

When reaching $P$ we can generically expect
to become maximally uncertain about the direction
of approach.
Looking at the expression (\ref{100.1}) for $\lambda_{i'}$
as glimpsed through the qmetric,
things go like if, when reaching $P$,
a direction,
that corresponding to a specific $|i'\rangle$,
is chosen at random.
The quantum behaviour would be in that
when approaching $P$ we become maximally uncertain about the
direction of approach, and in that the measurement of $\widehat{R}$
consists in extracting a random direction among those
corresponding to the $(D-1)$ eigenvectors $|i'\rangle$.

Another probe of $\widehat{R}$ at $P$ immediately after,
would correspond to a new random extraction of direction.
Still, quantum mechanics requires for the new measure that same
eigenvalue $\lambda_{i'}$.
This in itself poses a difficulty,
because, if the extraction is random,
we will in general expect $|i''\rangle \ne |i'\rangle$. 

At this stage,
one possibility  
might be to hypothesize, somehow by fiat, 
that the new pick of direction
is, for a system already probed, no longer at random.
Another one, figured in \cite{PesQ},
would be to allow
that the pick of direction is still at random
(according to what one would basically expect
from quantum mechanics),
but a specific mechanism, 
related to that the system has been
already probed, 
would prevent from
getting a result different from what already obtained. 
The mechanism would be in terms of a constraint
on $R_{ab}$, and then on the metric, and would involve 
endowing matter with the capability to affect curvature,
in such a way that matter could undo the variation
of the curvature 
one would get when 
going from $\lambda_{i'}$ to $\lambda_{i''} \ne \lambda_{i'}$. 
The system would keep this way for the Ricci scalar 
the value $\lambda_{i'}$ already obtained, 
and then as a matter of fact
would keep staying in the eigenstate $| i' \rangle$.
Interestingly, the mechanism takes the form of
field equations for the metric $g_{ab}$
thus possibly providing a quantum foundation for them.

Let us take stock for a moment.
What we have reached so far
is that, building on qmetric results,
it seems we can construct
a finite-dimensional Hilbert space
describing the quantum states of the gravitational field (the metric)
at an event $P$,
and we have given the explicit form of an operator
representing in this space the Ricci scalar.
This might be something interesting,
in that
there are
several proposed quantum descriptions of gravity,
but 
they in general do refer to {\it regions} of spacetime. 
The peculiarity here is instead to consider gravity 
in circumstances in which these regions
shrink (classically) to a point.

This result
might be expected,
at least according to some perspectives
as e.g. \cite{BaoCarrollSingh} we mentioned,
which do maintain that the number of gravitational
degrees of freedom in a local region ought to be finite,
and ask for what is the finite-dimensional Hilbert space
describing them. 
In this paper this is attempted
in the limit of these local regions shrinking to a point 
making use of minimum-length results.
This can be considered in a sense as a sort
of explicit implementation of the ideas of \cite{BaoCarrollSingh}
exploiting the fact that with spacetime endowed
with a minimum-length we have something which in the
small scale hints to a quantum structure. 

\section{Description in terms of component subsystems: entropy}

In the previous sections
we have considered a point $P$ in spacetime as a
coincidence event involving two physical systems:
a reference body (system $A$) and a (lightlike) test particle
(system $B$).
To accommodate the results of minimum-length metric
regarding the Ricci scalar,
we found then appropriate to describe the physics of 
the spacetime at $P$ in terms of
the composite system $A \otimes B$,
with states described by the (finite-dimensional)
Hilbert space $H$.
Our aim here,
is to try to gain some description of the states of $H$
in terms of the states of the component subsystems.
 
As one of the hallmarks of quantum theory,
we know that pure states of a composite system
can correspond to reduced states for the component systems
which turn out to be mixed.
Each state $|\psi^{AB}\rangle$ 
is meant to describe a situation 
in which the direction at $P$ is 
exactly given, and the state is accordingly pure.
Beside $|\psi^{AB}\rangle$,
we consider the states 
$\rho^A = \rm{tr}_B \, |\psi^{AB}\rangle$
of system $A$ and
$\rho^B = \rm{tr}_A \, |\psi^{AB}\rangle$
of $B$
we get when tracing out in each of the cases
the other system.

The physical situation we would consider 
is the following:
given circumstances in which 
the test particle reached $P$ along a definite direction,
we would like to describe the record of this in
system $A$ regardless of $B$ and vice-versa. 
The idea is that,
even if the direction at $P$ is exactly given for the coincidence,
namely for $A \otimes B$, this might no longer be true
for $A$ and $B$ taken separately.
In other words,
and focusing the discussion on $A$
(given its role as reference system),
it might not be true that,
after the test particle reached $P$,
in the reference frame at $P$ 
its arrival direction is known exactly.

We ask: 
does the existence of a minimum-length for spacetime
--property which, we have seen, might suggest a quantum description
for the system of the coincidence at $P$--
allow to make definite statements also regarding the states 
of the reference body $A$?    
To address this question let us define more precisely
what may be meant, from an operational point of view,
that the system $A$ finds the lightlike test particle
along some direction $\vec{k}$
and thus with tangent to the worldline 
$l^a = (k, \vec{k})$ ($k \equiv | \vec{k} |$). 

Imagine that the measurement of direction
is done say from the track left by the test particle
in $A$, taken this as some kind of spherical detector,
when going to
the coincidence limit $P'\to P$, i.e. $\lambda\to 0$,
with $P'$ the point at which the test particle is at
$\lambda$, which is the affine parameterization 
of geodesic such that $l^a = dx^a/d\lambda$ 
($x^a$ local coordinates) and 
$P'|_{\lambda=0} = P$.
Assume that $A$ is characterized by some angular resolution
and that this is captured in terms of some small solid angle
$\Omega_{(D-2)}$ along any given spatial direction
(the same in any given direction).

According to $A$, geodesics are straight lines 
in its Lorentz frame.
At some assigned $\lambda$,
$A$ takes samples
in a small solid angle $\Omega_{(D-2)}$
in any direction;
corresponding to the direction at which the signal results maximized
(maximum number of hits in the small solid angle 
in that direction)
$A$ takes every straight line from $P$
to points in the area $\Omega_{(D-2)} \lambda^{D-2}$
at $\lambda$ as a geodesics 
possibly describing
the track of the test particle;
this is the estimate $A$ gives of the
actual geodesic of the test particle,
based on the hits at $\lambda$
and on angular resolution $\Omega_{(D-2)}$.
When shrinking $\Omega_{(D-2)}$
(ideally even approaching 0) it obtains a better estimate,
and when doing this with 
$\lambda \to 0$
$A$ finally gets its measurement of the
arrival direction at $P$.

Beside the area $\Omega_{(D-2)} \lambda^{D-2}$,
we consider
the ($(D-2)$-dim)
area $a$ transverse to the direction of motion
of the test particle at any given $\lambda$
caught by the set of
geodesics from $P$
in the small solid angle $\Omega_{(D-2)}$
according to the actual metric
(at this stage, still unknown to $A$); 
these geodesics define $a = a(\lambda)$ 
along the
particle trajectory down to $P$.

What we would like to point out is that
whenever it happens that

\begin{eqnarray}\label{103.1}
a < \Omega_{(D-2)} \lambda^{D-2},
\end{eqnarray}
as we expect for curved spacetime
assuming null convergence condition holds,
this signals 
that in a measurement by $A$ characterized
by the solid angle $\Omega_{(D-2)}$
there is presence of spurious geodesics,
namely of geodesics which $A$ takes
as geodesics that trustworthy represent
the direction of the test particle within $\Omega_{(D-2)}$
but that actually do not.
This can be regarded as
the presence of a probability $p \ne 0$
that the geodesics taken by $A$ as representatives
within $\Omega_{(D-2)}$ of the true geodesic
are in reality not reliable for this.

We can try to define more precisely $p$ as follows.
We know the van Vleck
determinant $\Delta = \Delta(P', P)$ 
\cite{vVl, Mor, DeWA, DeWB} (see also \cite{Xen, VisA, PPV})
is the ratio 
of density of geodesics from $P$ at $P'$ between the actual spactime
under scrutiny and what would give the flat case \cite{VisA}.
Indeed, from the relation 

\begin{eqnarray}\label{78.4}
\theta 
=
\frac{D-2}{\lambda} - \frac{d}{d\lambda} \ln\Delta,
\end{eqnarray} 
where $\theta = \nabla_b l^b =
\frac{1}{a} \, \frac{d a}{d\lambda}$
is the expansion of the congruence,
we get
\begin{eqnarray}\label{78.7}
\ln \frac{a}{\lambda^{D-2}}
=
- \ln\Delta + C
\end{eqnarray}
with $C$ a constant,
which the consideration of flat case identifies 
as 
$C = \ln\Omega_{(D-2)}$.
This is

\begin{eqnarray}\label{78.11}
\frac{a}{\Omega_{(D-2)} \lambda^{D-2}}
=
\frac{1}{\Delta}.
\end{eqnarray}

We thus interpret
$a < \Omega_{(D-2)} \lambda^{D-2}$
as the presence of a probability

\begin{eqnarray}\label{79.1}
p(\lambda)
&\equiv&
\frac{\Omega_{(D-2)} \lambda^{D-2} - a}
{\Omega_{(D-2)} \lambda^{D-2}}
\nonumber \\
&=&
1 - \frac{1}{\Delta} > 0
\end{eqnarray}
for the geodesics within $\Omega_{(D-2)}$
to be mistakenly taken as a guess to the actual geodesic
of the test particle.
Clearly, all this makes sense as far as there is no caustic
along $\gamma$ in the interval we are considering; 
we can think this is always satisfied
provided we take $\lambda$ small enough.  

Notice that $p$ turns out to be independent
of $\Omega_{(D-2)}$.
This implies that
if circumstances are such that $p \ne 0$,
i.e. we are mistakenly guessing to some extent,
this is something which
cannot be cured on improving in angular resolution
(i.e. on taking a vanishing $\Omega_{(D-2)}$).

To characterize things at $P$,
we have to consider the $\lambda\to 0$ limit.
From the expansion \cite{Xen}

\begin{eqnarray}\label{vV}
\Delta^{1/2}(P', P) 
=
1 + \frac{1}{12} \, \lambda^2 R_{ab} l^a l^b 
+ \lambda^2 {\cal{O}}\bigg(\frac{\lambda}{L_R} \, R_{ab} l^a l^b\bigg)
\end{eqnarray}
(with the expression for the magnitude of higher
order terms taken from \cite{PesQ}),
where $R_{ab} l^a l^b$ is evaluated at $P$ and, we recall,
$L_R \equiv 1/\sqrt{R_{ab} l^a l^b}$
is the length scale proper, for the given $l^a$,
of the assigned curvature at $P$,
for $\lambda$ small we get

\begin{eqnarray}\label{vV_expansion}
p 
=
\frac{1}{6} \, \lambda^2 R_{ab} l^a l^b 
+ \lambda^2 {\cal{O}}\bigg(\frac{\lambda}{L_R} \, R_{ab} l^a l^b\bigg),
\end{eqnarray}
which, we see, gives $p\to 0$ when $\lambda\to 0$.
That is, there is no ineliminable probability
to be mistakenly guessing the actual geodesic
of the test particle starting from the track.

We can now proceed to inspect
what happens to this in a spacetime endowed with
a limit length. 
From the mere fact that
such a spacetime foresees the existence
of a non-vanishing area orthogonal
to the separation in the limit
of coincidence $P'\to P$ between the two points
(\cite{Pad06}; \cite{PesM, PesN, ChaD} for null geodesics),
we can expect that something deeply different
is going on in this case.

The minimum-length metric $q_{ab}(P', P)$ with base at $P$ and for
$P'$ null separated from $P$ is \cite{PesN} 
\begin{eqnarray}
\nonumber
  q_{ab} = {\cal{A}} \, g_{ab} +
  \Big({\cal{A}} - \frac{1}{\alpha}\Big) \,
  (l_a n_b + n_a l_b),
\end{eqnarray}
where
$l^a$ is the tangent to the null geodesic $\gamma$ connecting
$P$ and $P'$ and $n^a$ null is 
$n^a = V^a - \frac{1}{2} l^a$
with $V^a$ the velocity of the observer at $P$
(i.e. of the reference system $A$) parallel transported
along the geodesic. All these vectors are meant as considered
at $P'$.
The quantities
$\alpha$ and $\cal{A}$
are functions of $\lambda$
given by

\begin{eqnarray}\label{alphanull}
\nonumber
  \alpha =
  \frac{1}{d\tilde\lambda/d\lambda}
\end{eqnarray}
and

\begin{eqnarray}\label{Anull}
\nonumber
  {\cal{A}} = \frac{\tilde\lambda^2}{\lambda^2} \,
  \bigg(\frac{\Delta}{\tilde\Delta}\bigg)^{\frac{2}{D-2}},
\end{eqnarray}
where $\tilde\lambda$ is the qmetric-affine parameterization
of $\gamma$ expressing the distance along
$\gamma$ from $P$ as measured by the observer with velocity $V^a$,
with $\tilde\lambda \to L_0$ in the coincidence limit $P'\to P$. 
Here
$
\tilde\Delta(P', P) \equiv \Delta({\tilde P'}, P),
$
where $\tilde P'$ is that point on $\gamma$ (on the same side of $P'$)
which has $\lambda({\tilde P'}, P) = \tilde\lambda$.

In the qmetric,
the expansion $\theta_{(q)}$ is 
\cite{PesN, ChaD}

\begin{eqnarray}\label{rayqab2_31_3}
  \theta_{(q)}
  &=&
  \nabla_a^{(q)} \, l^a_{(q)}
  \nonumber \\
  &=&
  \alpha \,
  \Big[\theta + (D-2) \, \frac{d}{d\lambda} \ln \sqrt{\cal{A}}\Big]
  \nonumber \\
  &=&
  \frac{D-2}{\tilde\lambda} - \frac{d}{d\tilde\lambda} \ln \tilde\Delta.
\end{eqnarray}
Here,
$l^a_{(q)} = (d/d\tilde\lambda)^a = \alpha \, l^a$ is the tangent
to the geodesic at $P'$ according to the
qmetric-affine parameterization $\tilde\lambda$
and
the qmetric covariant derivative has the expression
$
\nabla_b^{(q)} \, v^a_{(q)}
=
\partial_b \, v^a_{(q)}
+ ({\Gamma^a}_{bc})_{(q)} \, v^c_{(q)}
$
for any qmetric vector $v^a_{(q)}$,
with the qmetric connection given by
$
({\Gamma^a}_{bc})_{(q)}
=
\frac{1}{2} q^{ad}
(-\nabla_d q_{bc} + 2 \nabla_{\left(b\right.}q_{\left.c\right)d}) + {\Gamma^a}_{bc}
$
\cite{KotG}
where
$q^{ab}$ (from $q^{ac} q_{cb} = \delta^a_b$) 
is 
$
q^{ab} 
=
\frac{1}{\cal{A}} \, g^{ab} +
\big(\frac{1}{\cal{A}} - \alpha\big) (l^a n^b + n^a l^b)
$. 
 
From
$
\theta_{(q)}
=
\frac{1}{a_{(q)}} \,
\frac{d a_{(q)}}{d \tilde\lambda}
$
(where $a_{(q)}$ is the $(D-2)$-dim transverse area
according to the qmetric),
following the same steps as above
with $\tilde\lambda$ and $a_{(q)}$
replacing respectively $\lambda$ and $a$, 
equation (\ref{rayqab2_31_3}) gives

\begin{eqnarray}\label{82.2}
\frac{a_{(q)}}{\Omega_{(D-2)} \tilde\lambda^{D-2}}
=
\frac{1}{\tilde\Delta},
\end{eqnarray}
analogous to equation (\ref{78.11}).

Then,
we interpret 
$a_{(q)} < \Omega_{(D-2)} \tilde\lambda^{D-2}$
as a probability

\begin{eqnarray}\label{82.3}
p(\lambda)
&\equiv&
\frac{\Omega_{(D-2)} \tilde\lambda^{D-2} - a_{(q)}}
{\Omega_{(D-2)} \tilde\lambda^{D-2}}
\nonumber \\
&=&
1 - \frac{1}{\tilde\Delta} > 0
\end{eqnarray}
to be mislead in taking
the geodesics within $\Omega_{(D-2)}$
as a guess to the actual geodesic
of the test particle.

Using (\ref{vV_expansion}),
for $\tilde\lambda \ll L_R$ 
(i.e. we are assuming to be close to the coincidence,
and that curvature is not too big
and we can have $L_0 \ll L_R$)
this gives

\begin{eqnarray}
p(\lambda)
=
\frac{1}{6} \, \tilde\lambda^2 R_{ab} l^a l^b 
+ \tilde\lambda^2 \, {\cal{O}}\bigg(\frac{\tilde\lambda}{L_R} \, 
R_{ab} l^a l^b\bigg).
\end{eqnarray}
At $P$, namely in the $\lambda\to 0$ limit,
we see that

\begin{eqnarray}\label{82.6}
p_0
&\equiv&
\lim_{\lambda\to 0} p(\lambda)
\nonumber \\
&=&
1 - \frac{1}{\Delta_0}
\nonumber \\
&=&
\frac{1}{6} \, {L_0}^2 \, R_{ab} l^a l^b 
+ {L_0}^2 \, {\cal{O}}\bigg(\frac{L_0}{L_R} \,
R_{ab} l^a l^b\bigg)
\nonumber \\
&\ne& 0,
\end{eqnarray}
where
$\Delta_0 \equiv
\tilde\Delta_{|\tilde\lambda = L_0}$.
Correspondingly,
\begin{eqnarray}\label{104.1}
a_{(q)}
&=&
\big(1 - p(\lambda)\big) \, \Omega_{(D-2)} \, \tilde\lambda^{D-2}
\nonumber \\
&=&
\frac{1}{\tilde\Delta} \, \Omega_{(D-2)} \, \tilde\lambda^{D-2}
\buildrel\lambda \to 0\over\longrightarrow
\frac{1}{\Delta_0} \, \Omega_{(D-2)} \, {L_0}^{D-2}
= a_0,
\end{eqnarray}
with
$a_0$ the limit $(D-2)$-dim orthogonal area 
\cite{PesM, PesN, ChaD} 
in the solid angle $\Omega_{(D-2)}$. 

We see,
equation (\ref{82.6}) may be interpreted
as showing that, 
in a spacetime with a minimum-length, 
the reference body $A$ has
an ineliminable
probability $p_0 \ne 0$ to be mistakenly guessing
the actual geodesic at the coincidence $P$.
We can understand this as follows.
Let us take a direction $\vec{k}$ as exactly known for the coincidence,
and thus described by the pure state $|\psi^{AB}\rangle = |\vec{k}\rangle$
of $A \otimes B$ associated to the null tangent $l^a = (k, \vec{k})$.
If $A$ takes a measurement at $P$ along this direction $\vec{k}$,
the outcome is not deterministic, 
even were the measurement performed
with infinite accuracy.
Indeed,
at ideal experimental conditions
still there is a probability $p_0$ that the system
is found in a state $|\alpha'\rangle \ne |\alpha\rangle$,
where 
$|\alpha\rangle$ is the (pure) state of $A$
with direction the assigned $\vec{k}$
and
$|\alpha'\rangle$ is the outcome pure state corresponding
to the ideal measurement of direction performed by $A$.
In other words,
the state $\rho^A$ of $A$ corresponding to $|\psi^{AB}\rangle$
would be, prior to the measurement, 
the mixed state

\begin{eqnarray}\label{104.2}
\rho^A
&=&
\rm{tr}_B |\psi^{AB}\rangle
\nonumber \\
&=&
(1 - p_0) \, |\alpha\rangle\langle \alpha |
+ p_0 \, \rho'^{A},
\end{eqnarray}
where
the density matrix $\rho'^{A}$ on the Hilbert space of $A$
has support
in the vector space orthogonal to $|\alpha\rangle$.

Summing up,
this result can be expressed as follows.
The observer $A$ is supposed to measure the arrival direction 
of an incoming photon $B$ nominally coming along 
some (null) direction $l^a$
($p\to P$ along geodesic with tangent $l^a$ at $P$).
Because of the existence of a finite limit area
transverse to $l^a$, there is an unavoidable 
(that is, present even assuming perfect resolution for the 
measuring apparatus)
nonzero probability $p_0$ that $A$ is mistaken in measuring
the actual arrival direction of photon $B$
arriving along $l^a$.
The observer has thus a probability $(1-p_0)$ to be 
correct in guessing
the arrival direction
and a probability $p_0$ to be not.  

On the basis of this we can do a little step more
in trying to understand the meaning of a $p_0 \ne 0$.
We ask what is the average gain of information $G$ by $A$
in guessing with perfect resolution the arrival direction of the photon
nominally arriving along $l^a$.
We know it is given by
the probability of correct guessing times 
the info associated to that guessing
+
the probability of incorrect guessing times 
the info associated to that other guessing
(cf. \cite{NielsenChuang}).
The infos here are log(probabilities) as we require that the info gained
in the occurrence of two independent events is the sum of the infos
of each event taken alone. We want moreover that the less probable is
an outcome, the greater is the gain of information we have 
in actually getting it.  
Normalizing to have $G = \ln{n}$ in case of $n$ different,
equiprobable outcomes,
we have  \cite{NielsenChuang}
$G = (1-p_0)\ln(1/(1-p_0)) +  p_0 \ln(1/p_0)$,
which is the Shannon entropy ${\cal H}(1-p_0, p_0)$ (in base $e$)
of the two-outcome probability distribution $(1-p_0, p_0)$.

A meaning for $p_0 = p_0(l^a)$ can be drawn
from the first term $I = I(l^a)$,
expressing the average gain of info in correct guessing,
of the just given expression for $G$,
\begin{eqnarray}\label{86.1}
I(l^a)
=
(1 - p_0) \, \ln \frac{1}{1 - p_0}
\end{eqnarray}
(this same term is present 
also in the expression of von Neumann's entropy $S(\rho^A)$
of state $\rho^A$ of (\ref{104.2}) (cf. \cite{NielsenChuang}): 
$
S(\rho^A) = 
{\cal H}(1-p_0, p_0)+(1-p_0) S(|\alpha\rangle\langle \alpha|)+p_0 S(\rho'^A) 
= {\cal H}(1-p_0, p_0)+p_0 S(\rho'^A) 
= (1-p_0)\ln(1/(1-p_0))+p_0 \ln(1/p_0)+p_0 S(\rho'^A),
$
where second equality is from being 
$|\alpha\rangle\langle \alpha|$ pure). 
Indeed,
in the $p_0 \ll 1$ limit
it reduces to

\begin{eqnarray}\label{86.2}
I(l^a)
&=&
(1 - p_0) \, \big(p_0 + {\cal{O}}({p_0}^2)\big)
\nonumber \\
&=&
p_0 + {\cal{O}}({p_0}^2)
\nonumber \\
&=&
\frac{1}{6} \, {L_0}^2 \, R_{ab} l^a l^b 
+ {L_0}^2 \, {\cal{O}}\bigg(\frac{L_0}{L_R} \,
R_{ab} l^a l^b\bigg),
\end{eqnarray}
where in the last equality we used the explicit expression of $p_0$ 
from (\ref{82.6}).

We see $p_0 \simeq I(l^a)$
for $p_0$ small,
and thus $p_0$ has the meaning 
of average gain of information by $A$
in correct guessing
of arrival direction 
(or correct guessing of state $|\alpha\rangle$
in von Neumann entropy's description).
Point is that with perfect resolution 
the observer $A$ would be supposed to always have
correct guessing (thus with no gain in info when
finding the nominal value $l^a$), 
were not for the indeterminacy connected with the existence
of finite limit area.

The understanding of $p_0$ as a gain of information
and its explicit expression (\ref{82.6})
suggests it might be reconnected with horizon entropy. 
Together with the photon from direction $l^a$
let us consider a local Rindler horizon at event $P$ \cite{JacB}
(then with expansion and shear exactly vanishing at $P$
in addition to an identically vanishing twist 
from hypersurface-orthogonality)
with generator $l^a$
(clearly this horizon is not the null congruence emerging from $P$
we used all along the paper).
The variation $\delta a$
of area
of a small patch $a$ of horizon
at event $P$ can be written as

\begin{eqnarray}\label{delta a}
\delta a
=
\bigg(\int_{-\bar\lambda}^0 \theta \, d\lambda\bigg) a
=
\bigg(-\int_{-\bar\lambda}^0 \lambda \, R_{ab} l^a l^b \, d\lambda\bigg) a
=
\frac{\bar\lambda^2}{2} \, R_{ab} l^a l^b \, a,
\end{eqnarray}
where the second equality stems from Raychaudhuri equation
$
\frac{d\theta}{d\lambda} 
=
- \frac{1}{D-2} \, \theta^2 - \sigma^2 - R_{ab} l^a l^b
$
as applied to the horizon,
with the first and the second term in the r.h.s. of higher order
with respect to the last.
In these expressions, 
$\bar\lambda$
is the width of the small affine interval
associated to the 
crossing of the horizon
by the test particle; $R_{ab}$ is the Ricci tensor at $P$.
Based on field equations, 
this area variation is associated with a variation of
horizon entropy, which precise expression (Wald entropy \cite{Wald, IyerWald}) 
depends on the actual gravitational theory under consideration.

But, we see that the expression (\ref{delta a})
is very similar to expression (\ref{82.6}) for $p_0$,
also reported in (\ref{86.2}) where 
$p_0$ finds interpretation
as average gain of information of $A$ in correct guessing the arrival
direction of the photon.
This suggests to reconnect the variation of horizon area (\ref{delta a})
with the gain of information by the observer at a single elementary event,
this hinting to
horizon area possessing an information content 
prior and regardless of
any invoking of gravitational field equations
(in the paper we never resort to field equations,
thus in particular they do not enter in deriving Eq. (\ref{86.2})).
In \cite{JacB} the association 
between horizon entropy and area
is motivated outside gravity 
by quantum field theory arguments
(entanglement entropy between vacuum fluctuations 
just inside and just outside the horizon);   
here it is made on the basis of kind of operational
and quantum-information arguments in a limit-length spacetime.

Expression (\ref{86.2}) also matches the formula
for the number of gravitational degrees of freedom 
$
\ln n_g 
\propto
\big(1 - \frac{L_{Pl}^2}{2\pi} R_{ab} l^a l^b\big)
$
given in \cite{Pad20}
($L_{Pl}$ is Planck length
and $n_g$ is the density of quantum states of spacetime),
used in the statistical derivation
of gravitational field equations.
This is not surprising
since here as there the starting point is the existence
of a non-vanishing limit area orthogonal to the geodesic
when $P' \to P$ in a spacetime with qmetric,
and we borrow from that approach
the individuation as key quantity
the ratio actual area/anticipated flat space area
and its $\lambda \to 0$ limit,
first considered in \cite{Pad06}.
The difference is in that 
in \cite{Pad06, Pad20} this has been introduced
to capture the number of microscopic dofs
of the (quantum) spacetime;
here to try to operationally show the mixedness of the quantum state
associated to coincidence according to reference body $A$,
and the ensuing arising of entropy. 

The results presented in this section might be summarized as follows.
The quantity $R_{ab} \, l^a l^b$, ubiquitous in spacetime thermodynamics
and responsible for horizon entropy in Einstein's gravity,
has been here found to be connected 
to the presence of unavoidable mixedness ($p_0 \ne 0$)
in the states describing 
spacetime, according to any observer,
at the most elementary level of a single (classical) event,
thus to kind of intrinsic 
ineliminable blurring in any observer's vision,
due to (the nature of) gravity itself 
not to limitations
on observer's side.

\section{Conclusions}

What we did in the paper,
has been to try to explore further the idea
that the intriguing results \cite{Pad01, KotI}
(and \cite{PesP} for null separations)  
concerning the form of the Ricci scalar
in a spacetime endowed with a minimum length
somehow might allow to sneak a look at
a quantum structure for spacetime at a point.
We did so 
with the conviction that
those results,
far from being artifacts 
of a mathematical model,
come about as
direct expression of a sound (possible) physical request,
as it is 
that of (consistently) requiring the existence of a minimum length.

For this further investigation, 
we have chosen
kind of an operational angle.
This being in part motivated
by the belief that
a most convenient way
to proceed in physics
is to keep close contact
with experiment,
through use of (at least in principle) 
as-well-as-possible
operationally defined concepts.
It is prompted also by 
the present flourishing
of activity in operational approaches to gravity,
with concrete hope of a direct experimental test
of a non-classicality of gravity
foreseeable in a hopefully not-so-far future
(\cite{BosA, MarlettoVedral} and subsequent proposals).

The results and main conclusions can be summarized as follows.
Building on the findings of the qmetric in the coincidence limit 
$p\to P$ of two events and limit length $L_0\to 0$,
it seems natural and possible to associate to a classical event $P$
a finite-dimensional Hilbert space describing the structure
of the qmetric spacetime at $P$.
A Hilbert space $H$ at $P$ 
has been explicitly 
built
as well as an operator
version of the Ricci scalar;
this Hilbert space describes crossing events at $P$ 
between an observer (state space $A$) and a test particle (state space $B$)
with the latter geodesically approaching $P$ ($p\to P$);
$H$ can then be taken $A\otimes B$.
It has been found that to pure states 
of $A\otimes B$ (exactly defined crossings)
do correspond (reduced density operator) 
mixed states of $A$ even with perfect experimental
resolution on observer's side, 
this inherent mixedness arising from the area transverse 
to the geodesic remaining finite when $p\to P$;
related to this, it has been found that 
there is an ineliminable (that is, present even with
perfect angular resolution) probability $p_0$, explicitly computed, 
for the observer to incorrectly guess the arrival direction.
We have seen how
this $p_0 \ne 0$ can be reinterpreted as average gain of information 
of the observer
when guessing correctly. 
Because the expression for $p_0$ is essentially analogous 
to the area variation of a suitable local Rindler horizon at $P$,
we noticed how this
might be used to provide 
an operational motivation
for endowing a patch of horizon 
with an entropy related to the area,
and this 
regardless of field equations 
(horizon entropy is not based on use of field equations).
In other terms, the qmetric 
appears to give the means to (operationally) introduce gravitational 
degrees of freedom before field equations.  

We can consider our results
in the context of other approaches also aiming to explore 
the consequences 
of a finite $L_0$
for spacetime description. 
In {\cite{NSW}} for example (see also {\cite{Nicolini}})
an effective metric is introduced
coming as what one gets
if in the propagator the source is smeared  
on a scale length $L_0$.
It has been usefully exploited to compute
corrections to the metric of black hole solutions 
due to a limit $L_0$ (getting in particular singularity
free spacetimes).  
At variance with the qmetric, 
this is an ordinary metric 
(distance $\to 0$ in the coincidence limit between
any two events),
even if corrected for effects due to $L_0$.
What the qmetric approach adds to this
is a handle to investigate 
the microstructure of the spacetime one gets 
if the limit length $L_0$ is embodied directly
in spacetime
(distance $\to L_0$ in the coincidence limit).
The study of this microstructure is what we 
dealt with here.

What we have presented is
essentially the observation that 
a puzzling result in the qmetric 
might find a description in terms of a quantum structure 
for spacetime
at $P$, provided $P$ is considered operationally 
as a coincidence event.
No emphasis has been put on the 
direct experimental scrutiny of this phenomenon
since as mentioned this seems hard to achieve, at least
as long as no signs of a limit-length $L_0 \ne 0$
are found at colliders.   
We may speculate however
that what described
might have if true a significance
with consequences in a sense also at low-energy lab scales.
The described quantum structure brings indeed with it 
that spacetime at a point $P$ might be considered as a superposition
of classical geometries at $P$.
The limit length $L_0$ sets the scale at which the 
quantum features are expected to unavoidably show up,
where `unavoidably' means for spacetime generic.
We can expect that for specific spacetimes,
the quantum features might show up 
at much larger scales.
For example we can consider the spacetime sourced 
by a delocalized particle:
the point is that if superpositions
are unavoidably found at scale $L_0$
for undelocalized sources,
it is not unreasonable to expect that they might appear, 
at larger scales,
also due to the delocalization of the source.
This spacetime might then consist
of a superposition 
of the classical spacetimes
corresponding each to a superposed position 
of the particle,
and the related effects might be in principle detectable also
at low-energy lab scales 
(as in the proposals 
\cite{BosA, MarlettoVedral}).

{\it Acknowledgments.} This work was
supported in part by INFN grant FLaG.


\end{document}